\documentclass[journal]{IEEEtran}
\usepackage[noadjust]{cite}
\usepackage{multirow} 
\usepackage{algpseudocode}
\usepackage{algorithm}
\usepackage{rotating}
\usepackage{fancyvrb}
\usepackage{kantlipsum} 
\usepackage{commath}
\allowdisplaybreaks
\usepackage{mathtools}  
\usepackage{verbatim}
\usepackage{xr-hyper} 
\usepackage{enumitem}

\usepackage{listings}

\usepackage{epstopdf}
\usepackage{wrapfig}
\usepackage{latexsym}
\usepackage{amssymb}
\usepackage{amsthm}
\usepackage{amsfonts}
\usepackage{amsmath} 
\usepackage{graphicx}
\usepackage{latexsym}
\usepackage{booktabs}
\usepackage{support-caption} 
\usepackage{subcaption} 
\usepackage{xspace}

\usepackage[normalem]{ulem} 

\usepackage[T1]{fontenc}
\usepackage{listings}
\usepackage{protobuf/lang}  
\usepackage{protobuf/style} 

\usepackage{placeins}

\ifCLASSINFOpdf
\else
\fi
\newcommand{\dq}[1]{``#1''}

\newif\ifcommentson
\commentsontrue

\newif\ifextended
\newif\ifshortver

\shortvertrue

\newif\ifrevision

\begin{document}


\title{SRv6-PM: Performance Monitoring of SRv6 Networks with a Cloud-Native Architecture}

\author{Pierpaolo Loreti, Andrea Mayer, Paolo Lungaroni, Francesco Lombardo, Carmine Scarpitta, Giulio Sidoretti\\
Lorenzo Bracciale, Marco Ferrari, Stefano Salsano, Ahmed Abdelsalam, Rakesh Gandhi, Clarence Filsfils
\IEEEcompsocitemizethanks{\protect
\IEEEcompsocthanksitem P. Loreti, A. Mayer, F. Lombardo, L. Bracciale, S. Salsano are with the Department of Electronic Engineering at the University of Rome Tor Vergata and with the Consorzio Nazionale Interuniversitario per le Telecomunicazioni (CNIT) - Rome, Italy, E-mail: \{stefano.salsano, pierpaolo.loreti, andrea.mayer\}@uniroma2.it.
\IEEEcompsocthanksitem P. Lungaroni, G. Sidoretti are with the Consorzio Nazionale Interuniversitario per le Telecomunicazioni (CNIT) - Rome, Italy.
\IEEEcompsocthanksitem C. Scarpitta is with the Department of Electronic Engineering at the University of Rome Tor Vergata and the Consortium GARR - Rome, Italy.
\IEEEcompsocthanksitem M. Ferrari is with the Department of Electronic Engineering at the University of Rome Tor Vergata - Rome, Italy.
\IEEEcompsocthanksitem A. Abdelsalam, R. Gandhi, C. Filsfils are with Cisco Systems, E-mail: \{ahabdels, rgandhi, cfilsfil\}@cisco.com
}
\vspace{2ex}
\textbf{\\Submitted paper - Version 1 - July 2020}
\vspace{-3ex}
}

%
%

\markboth{Submitted to IEEE TNSM}%
{Shell \MakeLowercase{\textit{et al.}}: Bare Demo of IEEEtran.cls for IEEE Journals}
%



\maketitle

\begin{abstract}

Segment Routing over IPv6 (SRv6 in short) is a networking solution for IP backbones and datacenters. The SRv6 standardization, implementation and research are actively progressing and SRv6 has already been adopted in a number of large scale network deployments. Effective Performance Monitoring (PM) solutions for SRv6 networks are strongly needed. The design, implementation and deployment of such PM solutions span the different planes of a networking architecture: Performance Measurements data (packet loss and delay) needs to be measured (in the Data Plane), the monitored nodes needs to be controlled (in the Control Plane), the measured data needs to be collected (in the Control/Management Planes), then the Data must be processed and stored, using Big-Data processing solutions. 

We focus on Loss Monitoring, by considering a solution capable of tracking single packet loss events in near-real time (e.g. with a delay in the order of 20 seconds).

We describe SRv6-PM, a solution for Performance Monitoring of SRv6 networks. SRv6-PM features a cloud-native architecture for the SDN-based control of Linux routers and for ingestion, processing, storage and visualization of PM data. In the Data Plane, SRv6-PM includes efficient building blocks for packet loss evaluation (e.g. the packet counting components) in a Linux router.

SRv6-PM is released as open source. Not only we provide a reproducible environment for PM experiments, but we also offer a re-usable and extensible cloud-native platform that can be automatically deployed in different environments, from a single host to multiple servers on private/public clouds. 

\end{abstract}


%
\IEEEpeerreviewmaketitle

\section{Introduction}

Novel paradigms such as Software Defined Networking, Network Function Virtualization and Network Virtualization can increase flexibility and reliability of high speed networks if supported by effective tools and systems able to monitor the health of the infrastructure and its offered performances continuously and accurately.  Classical monitoring tools and protocols have been evolving to meet the new challenges of softwarized networks, and several new solutions \cite{akyildiz2014roadmap} and protocols have been presented, standardized and effectively applied \cite{tsai2018network}. In recent years, Segment Routing (SR) networking technology has come to prominence mainly to cope with the needs of 5G networks or geographically distributed large scale data centers.  SR is based on loose source routing: a list of \textit{segments} can be included in the packet headers. The segments can represent both topological way-points (nodes to be crossed along the path towards the destination) and specific operations on the packet to be performed in a node. Examples of such operations include: encapsulation and de-capsulation, lookup into a specific routing table, forwarding over a specified output link, Operation and Maintenance (OAM) operations such as time-stamping a packet. More generally, arbitrarily complex behaviors can be associated to a segment included in a packet.

Segment Routing (SR) architecture can be deployed on the IPv6 data plane, \cite{rfc8402,filsfils2015segment} and its implementation can be beneficial also to support a smooth network migration towards IPv6. In Segment Routing for IPv6 (SRv6 in short), the segments are represented by IPv6 addresses and are carried in an IPv6 \textit{Extension Header} called \textit{Segment Routing Header} (SRH) \cite{ietf-6man-segment-routing-header}. The IPv6 address representing a segment is called SID (Segment ID). According to the SRv6 \textit{Network Programming} concept \cite{id-srv6-network-prog}, the list of segments (SID List) can be seen as a \dq{packet processing program}, whose operations will be executed in different network nodes. The SRv6 Network Programming model offers an unprecedented flexibility to address the complex needs of transport networks in different contexts. With the SRv6 Network Programming model, it is possible to support valuable services and features such as layer 3 and layer 2 VPNs, Traffic Engineering, fast rerouting, etc. A tutorial on SRv6 technology can be found in \cite{ventre2019survey}, \cite{tian-spring-srv6-deployment-consideration-02}. The standardization activities for SRv6 are actively progressing in different IETF Working Groups, among which the SPRING (Source Packet Routing In NetworkinG) WG is taking a leading role. Recently, several large scale deployments in operator networks have been disclosed, as reported in \cite{matsushima-spring-srv6-deployment}.

Performance Monitoring (PM) is a fundamental function to be performed in softwarized networks. It allows operators to detect issues in the QoS parameters of active flows that may require immediate actions and to collect information that can be used for the offline optimization of the network.  SRv6 PM solutions can be analysed considering two separate subsystems: i) modules and protocols integrated into the data plane to measure and collect data related to nodes and individual traffic flows; ii) data management infrastructures specifically designed to store, organize and analyze monitoring data collected in the network. 
Regarding the Performance Monitoring data plane subsystem for SRv6, two Internet Drafts have been proposed and are currently under discussion in the IETF SPRING WG. These drafts rely on existing methodologies for performance measurement in general IP and MPLS networks. They propose the extension of such methodologies to the SRv6 PM case. Both proposed solutions focus on system architecture and protocol specification, but the actual system implementation and integration in the network data plane must still be defined and validated in the field. 

Moreover, large networks usually consist of hundreds of nodes generating a huge amount of data, and a new class of problems arises when considering the required storage and elaboration capacity. This scenario calls for the integration of a Cloud Native Big Data solution with the ability to support common management tasks. Several Network management solutions comprise a cloud ready architecture such as Nagios \cite{nagios}, and in there are also few proposals that specifically tackle the requirements of performance monitoring systems such as \cite{queiroz2019approach}.

In this paper we study and discuss the proposed SRv6 PM approaches, considering both data plane and control plane aspects. Moreover we devise a complete open architecture for performance monitoring of SRv6 networks, including the data plane part and a cloud native big data ready management part based on available open source technologies. To validate the proposed architecture we implemented an SDN accurate Per-Flow Packet Loss Measurement (PF-PLM) solution for SRv6 flows based on Linux kernel networking. The proposed solution integrates the \dq{alternate marking} method described in RFC 8321 \cite{rfc8321} and provides an accurate estimation of flow level packet loss (it achieves single packet loss granularity). We also devised a cloud native architecture using the available open source tools specifically designed to collect topological data and time series related to the various SRv6 traffic flows.

In detail the main novel contributions of this work are:
\begin{itemize}
    \item definition of an architectural solution for the Performance Monitoring of SRv6 traffic flows compliant with available standards and internet drafts; 
    \item definition of a gRPC Southbound interface for the PM;
    \item integration of an alternate marking method for Loss Monitoring in SRv6 networks; 
    \item design and implementation of a cloud-native architecture for PM based on open source tools;
    \item implementation of a Per-Flow Packet Loss Measurement solution complaint with the TWAMP extension proposed in \cite{gandhi-spring-twamp-srpm-08} on a Linux router; 
    \item implementation of eBPF based packet counting components for Linux SRv6 networking;
    \item evaluation of performance degradation introduced by the Loss Monitoring (packet counting) solution and comparison with previous implementation based on IPSet.
\end{itemize}

The paper is organised as follows: in section \ref{sec:soastd} we present the related work and the relevant standards on Performance Monitoring, as well as the proposed solution for SRv6; in section \ref{sec:archsol} we detail the proposed architecture for PM in SRv6 and in section \ref{sec:losssol} the corresponding casting to the PF-PLM solution. In section \ref{sec:impl} we describe several implementation details of the Linux Kernel components and the deployment of the cloud native architecture. In section \ref{sec:testbeds} we presents the implemented test beds and in \ref{sec:perf} we validate the proposed architecture and evaluate the performance of the data plane components. Finally we draw the conclusions.

\section{Performance Monitoring solutions and standardization} \label{sec:soastd}
In this section we begin by introducing the related work on Performance Monitoring for Softwarized Networks.  We then provide an overview on the related relevant standards for IP and MPLS networks before finally discussing the solutions for SRv6 proposed for IETF standardization.

\subsection{Performance Monitoring in Softwarized Networks}
There are many commercial and open source solutions for network performance monitoring. Some of them rely on generic tools that, among other features, include the monitoring of network devices. Two notable examples are Nagios \cite{nagios} and Zabbix \cite{zabbix}. Other solutions are based on tools developed to monitor cloud environments, such as Ceilometer \cite{ceilometer} for example, which is adopted in Open Stack deployments. In the SDN, OpenFlow protocol is the industry standard de facto, and its interface allows controllers to obtain from nodes numerous statistics about the flows that the device is managing \cite{openflow}. Several studies have proposed solutions for OpenFlow networks. For example in \cite{van2014opennetmon}  OpenNetMon proposes a framework to measure throughput, delay, and packet loss of traffic flows in OpenFlow networks. The work focuses on how to effectively collect data to provide controllers with a network-wide vision but limiting the additional computational load needed to obtain all measurements. Another solution for OpenFlow is proposed in \cite{santos2015network} where the cost of such measures is also discussed. A recent overview of the activities on Openflow traffic monitoring is provided in \cite{queiroz2019approach}.

\subsection{Active Monitoring in IP and MPLS networks}
Active measurements can be an effective solution to enable the monitoring of some performance metrics such as loss and one-way or two-way delays following the so called fate sharing paradigm, according to which probe and data packets share the same network "fate". 
Several research works and standards have been proposed both for IP and MPLS networks, especially in the IETF framework. Among them, RFC 4656 \dq{The One-Way Active Measurement Protocol (OWAMP)} \cite{rfc4656} provides capabilities for the measurement of one-way performance metrics in IP networks such as one-way packet delay and one-way packet loss. RFC 5357 \dq{Two-Way Active Measurement Protocol (TWAMP)} \cite{rfc5357} introduces the capabilities for the measurements of two-way (i.e. round-trip) metrics. These specifications describe both the \textit{test protocol}, i.e. the format of the packets that are needed to collect and carry the measurement data and the \textit{control protocol} that can be used to setup test sessions and to retrieve measurement data. For example OWAMP defines two protocols: \dq{OWAMP-Test is used to exchange test packets between two measurement nodes} and \dq{OWAMP-Control is used to initiate, start, and stop test sessions and to fetch their results} (quoting \cite{rfc4656}). Note that in general there can be different ways to setup a test session: the same test protocol can be re-used with different control mechanisms. 

RFC 6374 \cite{rfc6374} specifies protocol mechanisms to enable the efficient and accurate measurement of performance metrics in MPLS networks. The protocols are called LM (Loss Measurement) and DM (Delay Measurement). We will refer to this solution as MPLS-PLDM (Packet Loss and Delay Measurements). In addition to loss and delay, MPLS-PLDM also considers how to measure throughput and delay variation with the LM and DM protocols. Differently from OWAMP/TWAMP, RFC 6374 does not rely on IP and TCP, and its protocols are streamlined for hardware processing. While OWAMP and TWAMP support the timestamp format of the Network Time Protocol (NTP) \cite{rfc5905},  MPLS-PLDM adds support for the timestamp format used in the IEEE 1588 Precision Time Protocol (PTP) \cite{IEEE1588}. There are several types of channels in MPLS networks over which loss and delay measurement may be conducted. Normally, PLDM query and response messages are sent over the MPLS Generic Associated Channel (G-ACh), which is described in detail in RFC 5586. RFC 7876 \cite{rfc7876} complements the RFC 6374 by describing how to send the PLDM response messages back to the querier node over UDP/IP instead of using the MPLS Generic Associated Channel.

\subsection{SRv6 Performance Monitoring}

The standardization activity regarding the Performance Monitoring of SRv6 networks is very active and two internet drafts have been published so far:
\begin{itemize}
    \item Performance Measurement Using UDP Path for Segment Routing Networks \cite{gandhi-spring-rfc6374-srpm-udp-03}
    \item Performance Measurement Using TWAMP Light for Segment Routing Networks \cite{gandhi-spring-twamp-srpm-08}
\end{itemize}
Both solutions provide the possibility of measuring delay and loss of a single SRv6 flow, characterized by a SID list. The data collection takes place with test UDP packets transmitted on the same measured path. The test UDP packets collect the one way or two way PM data and make them available to the node which requested the measurement.

The first one aims at extending and reusing the MPLS-PLDM work defined in RFC 6374 \cite{rfc6374} and RFC 7876 \cite{rfc7876} and specifies procedures for using a UDP path for sending in-band probe query and response messages for Delay and Loss performance measurement. Although the RFC 6374 applies only to MPLS Networks, the  specified procedures are applicable to SR-MPLS and SRv6 data planes for both links and end-to-end measurement for SR Policies. The draft introduces TLV to specify the Return Path and a TLV for traffic coloring. Another interesting characteristic of the proposal is the possibility to send the probe response directly to an external controller. Although the proposed solution seems consistent and mature (its standardization started in March 2018), the draft expired in March 2019.

The most active standardization path \cite{gandhi-spring-twamp-srpm-08}  promotes the adoption of a modified version of the TWAMP Light protocol defined in RFC 5357 Appendix I and its simplified extension Simple Two-way Active Measurement Protocol (STAMP) proposed for standardization in \cite{ietf-ippm-stamp-10}. These protocols lack support for Loss Measurement in traffic flows that required in SRv6 networks. Thus the draft uses procedures and messages defined in RFC 5357 for Delay Measurement (DM) and specifies new procedures and messages for Loss Measurement both for SR-MPLS and SRv6 data planes. We select this draft as reference procedures for the performance monitoring architecture presented in this work and implemented for evaluation.

Moreover both the solutions support the Alternate-Marking Method defined in RFC 8321 \cite{rfc8321} for accurate loss monitoring and that can be applied to IPv6 and SRv6 flows as specified in \cite{fz-6man-ipv6-alt-mark-09} and presented by \cite{mizrahi2019pm}. 
Indeed the presence of in flight packets makes it difficult to obtain an accurate evaluation the number of lost packet, as discussed in RFC 8321 \cite{rfc8321}. The proposed solution combines packet ``marking'' and packet counting to cope with this problem. We integrated this technique in our accurate loss monitoring solution, and a detailed description is provided in section \ref{sec:losssol}.

A preliminary version of the coloring and counting solution based on the Linux IPTable modules was studied and presented in \cite{loreti2020implementation}. In this work we overcome the identified limitations developing a new packet counting component based on the eBPF framework. Note that the previous work \cite{loreti2020implementation} did not include the definition of the cloud-native architecture and the SRv6-PM platform.

\section{Monitoring architecture}\label{sec:archsol}

\subsection{Performance Monitoring: Data and Control Planes}

\begin{figure}[t!]
    \centering
    \includegraphics[width=3in]{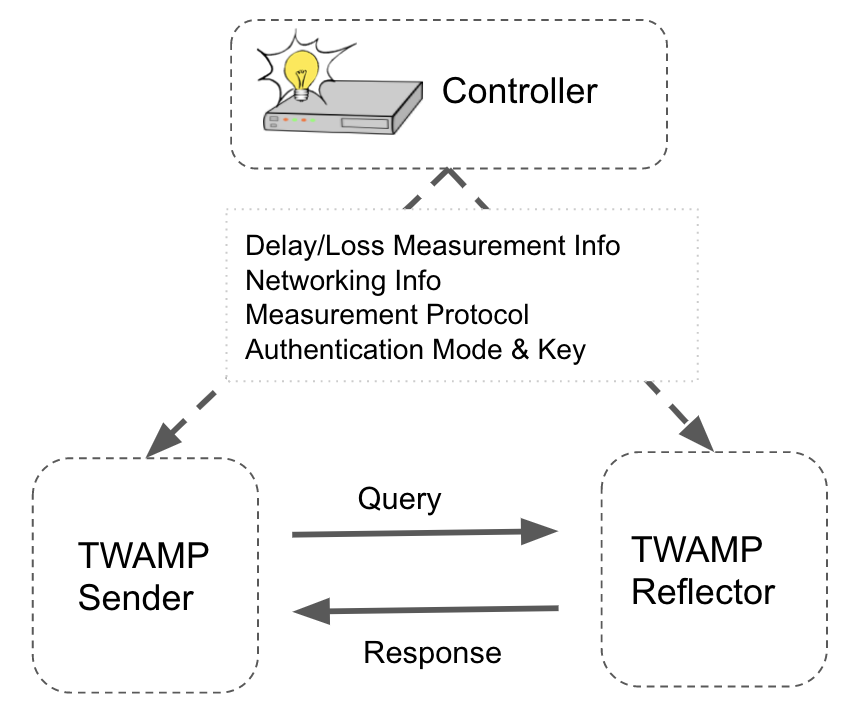}
    \caption{Architecture of Performance Monitoring Data and Control planes}
    \label{fig:dparch}
\end{figure}

\begin{figure*}[t!]
    \centering
    \includegraphics[width=\linewidth]{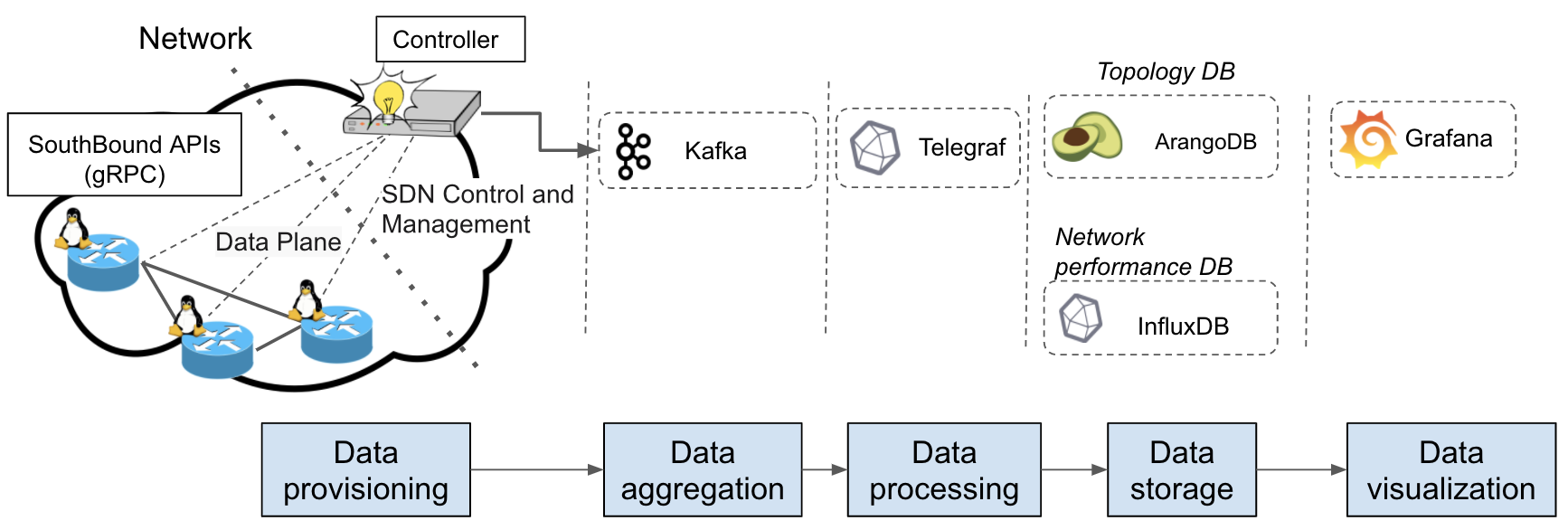}
    \caption{Architecture of the data monitoring and analytics}
    \label{fig:bigdataarch}
\end{figure*}


To monitor the QoS experienced by the SRv6 traffic flows, the controller needs to interact with the network routers and switches. The main operations are to start/stop the monitoring procedures on the selected flows, and then to collect the measurement data such as packet loss ratio and delay. Moreover, a data plane measurement protocol is needed among the monitored nodes. In our proposed architecture for SRv6 PM, we adopt the approach described in \cite{gandhi-spring-twamp-srpm-08}. It uses and extends the TWAMP Light protocol, which appears more suitable for the IPv6 data plane with respect to \cite{gandhi-spring-rfc6374-srpm-udp-03} which was conceived for an MPLS data plane. Fig \ref{fig:dparch} shows the reference architecture for the data plane monitoring protocol, including the interaction of the data plane protocol entities with the controller. Two entities are considered, named \textit{TWAMP Sender} and \textit{TWAMP Reflector} respectively. The \textit{TWAMP Sender} sends a probe \textit{Query} to the \textit{TWAMP Reflector}, which replies with a probe \textit{Response} message. The Query and the Response messages carry the performance monitoring information. The Controller shown in Fig \ref{fig:dparch} is used to provision the TWAMP sender, and is received with the proper configuration information to carry out the measurements to start/stop the experiment and to collect the results. The presence of the Controller eliminates the need for a \dq{control plane} monitoring protocol between the TWAMP sender and the TWAMP receiver: all control and management operations are performed by the controller.

Both the probe query and response messages are sent on the congruent path of the data traffic by the sender node. They are then used to measure the delay of an SRv6 traffic flow or to collect counters related to a specific flow. The controller needs to specify the type of measure that needs to be performed, the SID list of the monitored flow, the relevant networking information such as the UDP port and the destination address, the authentication mode and keys.

The controller interacts with the network nodes using an API, which needs to be carefully designed. We decided to extend the SRv6 Southbound API proposed in \cite{ventre2018sdn}, in order to allow full control of the SRv6 data plane and routing of a Linux node. In \cite{ventre2018sdn} the authors had prototyped, validated and compared three different implementations:  gRPC, REST, NETCONF. In this work we have adopted the gRPC solution for API implementation. The Performance Monitoring architecture proposed in \cite{gandhi-spring-twamp-srpm-08} does not explicitly provide for a data collection system, and therefore in this work we devised two complementary data collection solutions:
\begin{itemize}
    \item Data collection is operated directly by the Controller via the Southbound APIs provided by the nodes.
    \item The Controller pushes the Measurement data to the Cloud Native Data Infrastructure using the Publish-Subscribe paradigm. 
\end{itemize}
The used data collection solution is configured by the controller using the Southbound interface.

\subsection{Cloud Native Big Data Management}

The data retrieved from the measurement processes should be collected together with a rich set of companion data which describes the state of the network at that specific time as well as adding other contextual information. This information can be used both for real-time monitoring reasons and for offline analysis and optimization (e.g. for traffic engineering or network planning applications). In both cases, the data should be easily available for monitoring network status in real-time, and to provide long term insights using the more suitable and powerful tools (e.g. machine learning based processes, forecasting and general analytics).

The above requirements call for a dedicated Big Data infrastructure to take care of all the different phases of the data path, from the data ingestion to the visualization. In particular, we foresee the following steps of the data management pipeline:
\begin{itemize}
    \item \textbf{Data provisioning}: The controller collects Data from the network nodes through the (gRPC) Southbound APIs. Data includes measurement specific information as well as generic information on the network such as the current topology status.
    \item \textbf{Data aggregation}: A network can have multiple controllers and data sources, which need to be conveyed in the data storing systems. This data should be ingested through a scalable data ingestion pipeline. 
    \item \textbf{Data processing}: Data must be pre-processed before its storage. The goals of this processing can be multiple: from data adaptation to efficient data representation. For instance, it can be necessary to parse different input data formats such as logfile chunks to extract the metrics of interest. 
    \item \textbf{Data storage}: Dedicated big data storage must be able to accumulate data for both instant monitoring and future analysis (e.g., machine learning). Data representation is strictly correlated with the selected Database system. Graph data and time-series data can thus be stored on different data stores.
    \item \textbf{Data visualization}: Finally, a system for monitoring and analytics is needed to show the result of the analysis and to offer proper dashboards to traffic engineers.
\end{itemize}

In our architecture we have selected \textit{open source} tools for each of the above tasks, rather than using proprietary systems offered by technology providers or by cloud infrastructure providers. This has the benefit of avoiding vendor lock-in, but might result in a more complex integration given that the whole pipeline is not shipped as-a-whole by a single provider. For this reason, we believe that it is not enough to \dq{statically} define the reference architecture, but we also ship a fully integrated CI/CD (Continuous Integration / Continuous Development) environment, that will be described in section \ref{sec:cicd}.

The selected architecture is depicted in figure \ref{fig:bigdataarch}. We start from the so-called TIG stack (Telegraf, InfluxDB, Grafana)  and add services to cope with the specific requirements of our scenario. We briefly go through the components describing which task they accomplish before describing how they are dynamically correlated.

\begin{itemize}
    \item \textbf{Data aggregation $\rightarrow $ Apache Kafka}: Apache Kafka is a publish subscribe system designed for routing large-scale real-time feeds.  
    Kafka has an undoubtedly good reputation as a low latency and high throughput platform. With respect to other open source solutions such as Apache Storm or Apache Spark, Kafka offers less redundant operation inside the proposed pipeline. The simple decoupling pub-sub interface of producers and consumers allows an easy decoupling of the many controllers (and other data feeds) from the rest of the data pipeline. We used Kafka to collect information coming from one or more controllers, and to inject it into the next block of the chain (Data processing). As represented in figure \ref{fig:bigdataarch}, data are pushed from the controller to the Kafka sub-system: in general this architecture allows multiple controllers and, more generally, multiple data sources.
    Indeed, using Kafka we can easily extend our system by adding new data provisioning modules.
    \item \textbf{Data processing $\rightarrow $ Telegraf}: Telegraf is a server agent designed to collect metrics and pre-process data. Alternative solutions (e.g., logstash from the ELK stack) appear to be more deeply connected inside their related stack, and thus less suitable to work stand-alone in a different environment. Telegraf can be used in the data processing phase, but also in the data provisioning just by moving its software agent directly on the device we are going to monitor. In our case we used Telegraf to pre-process the data read from Kafka and coming from the controller moving to the Data storage step.
    \item \textbf{Data storage $\rightarrow $ InfluxDB and ArangoDB}: We selected two different Database systems, one for the time series data, and another one specialized to handle graph data for storing the topology. Both systems are NoSQL, scalable and well supported. As for the time series Database, InfluxDB provides a fully fledged set of features for visualization, alerts and triggers, as well as primitives for anomaly detection and machine learning. For graph data, Arango is a NoSQL database for big data with native support for graphs. We use it for storing the topology of the network which is provided directly by the controller.
    \item \textbf{Data visualization $\rightarrow $ Grafana}: Grafana is a monitoring and analytics tool that we use to show the performance collected inside InfluxDB on a monitoring realtime cockpit.  With respect to other systems (e.g. Kibana, Nagios) it can be highly customized (with Grafana Plugins), offers broad support and it is not part of a specific stack which could make its integration more difficult.
\end{itemize}

The practical deployment of the above described architectural stack into our open source testbed is described in section \ref{sec:cicd}.


\section{SRv6 Accurate Loss Measurement}\label{sec:losssol}

\begin{figure}[t!]
    \centering
    \includegraphics[width=0.48\textwidth]{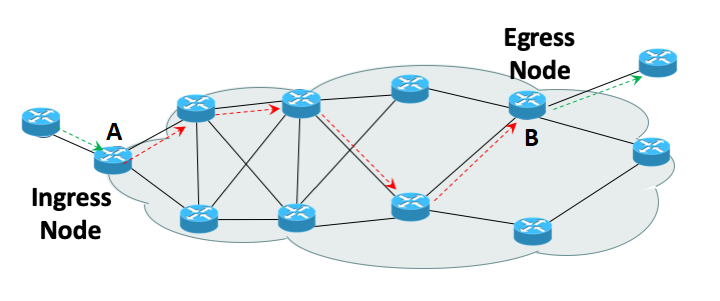}
    \caption{Reference SRv6 network scenario for Performance Monitoring}
    \label{fig:pm-data-plane}
\end{figure}

To validate the proposed architecture, we devised a full Per Flow Packet Loss Measurement (PF-PLM) solution, compliant with the draft \cite{gandhi-spring-twamp-srpm-08} and adopting the step detection the alternate marking technique for counting described in \cite{mizrahi2019pm}. We consider the reference network scenario depicted in Figure \ref{fig:pm-data-plane} comprising an SRv6 network domain. IP traffic arrives at an ingress edge node of the network where it can be classified and encapsulated in an SRv6 flow. In SRv6 terminology, an \textit{SR policy} is applied to the incoming packets. The SR policy corresponds to a \textit{SID List} that is included in the Segment Routing Header (SRH) of the outer IPv6 packet. The outer IPv6 packet crosses the network (according to its SID list) and arrives to the egress edge node where the inner IP packet is decapsulated (the outer IPv6 and SRH are removed). For example in the figure \ref{fig:pm-data-plane}, node $A$ acts as ingress node while node $B$ as egress node for the green packets. The ingress node $A$ applies the SR policy, i.e. it writes the \textit{SID list} into the SRH header.

\subsection{Packet Counting}

In order to perform loss measurement we need to implement packet counters associated to SRv6 flows, both in the ingress node and in the egress node. For our purposes, an SRv6 flow corresponds to an SR policy, i.e. to a SID List. We want to be able to explicitly activate the counting process for a set of flows (identified by their SID Lists).

In an ingress node this means to process all outgoing SRv6 packets and count the packets belonging to the set of monitored flows (by comparing the SID List of the outgoing packets with the SID Lists of the monitored flows). Likewise, in an egress node, this means to process all incoming SRv6 packets, check if the packets belong to the set of monitored flows and increment the counters as needed. 
These counting operations can have a high computing cost for a software router, so it is important to carefully design their implementation (see section \ref{sec:impl} to evaluate their impact on the processing performance).

\subsection{Traffic Coloring}

The comparison of the transmission counter with the receiver counter is simple to implement but requires that the two counters refer exactly to the same set of packets and since flows cannot be stopped it is difficult to get an accurate loss evaluation. If we want to achieve the granularity to accurately detect single packet loss events, while the flow is active, we need to properly consider the \dq{in flight} packets, e.g. the packets counted by the ingress node but not by the egress one. 

The solution proposed in the RFC 8321 \cite{rfc8321} is to virtually split packets into temporal countable blocks and by ``coloring'' the packets of the flows to be monitored with at least two different marks so that different consecutive blocks will have different colors. For example, a continuous block of packets of a flow is colored with color $R$ (e.g. for a configurable duration $T$), then the following block of packets is colored with color $B$ (again for a duration $T$), and so on. In this case two separate counters are needed both in the ingress node and in the egress node for packets with color $R$ and with color $B$. In general the this solution requires for each flow one counter per color per interface.

\begin{figure}[t!]
    \centering
    \includegraphics[width=0.48\textwidth]{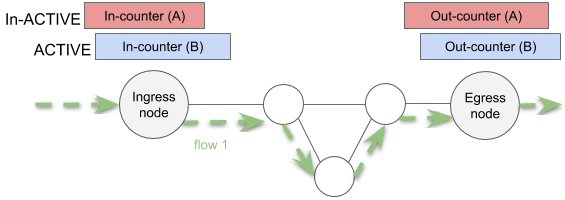}
    \caption{Alternate coloring method (RFC 8321)}
    \label{fig:alternate-coloring}
\end{figure}

To calculate the packet loss we can read the inactive counters. In the previous example, when the active counter is $B$, it is possible to safely read counters for color $R$ from the ingress and egress nodes, and evaluate their difference which exactly corresponds to the number of lost packet in the previous interval (see Fig.~\ref{fig:alternate-coloring}). 
Usually it is worth waiting some time to read the counters after the color switch in order to give time to in flight packets to arrive at the egress node. In our implementation we wait $T/2$ before reading the inactive counter.  The main drawbacks of this techniques are i) the large number of counter needed and ii) that the measurement frequency is limited by the block duration. 

RFC 8321 describe a generic method that can be applied to different networks. When the alternate marking method is applied to a specific network, the mechanism to color the packets must be specified. In \cite{fz-6man-ipv6-alt-mark-09} is reported a extended discussion on the possible alternatives for marking IPv6 packets. They consider solutions based on the IPv6 Extension Headers, IPv6 Address or Flow Label, but concludes that ``the only safe choice that can be standardized would be the use of a new TLV''. However, in the paper \cite{mizrahi2019pm} from the same authors, simpler techniques have been considered and discussed. 

In our architecture, since we are considering an SRv6 network we can leverage on the intrinsic network programming model and thus we considered two possible alternatives: i) modification of the DS field, previously known as IP Type of Service (TOS) field; ii) encoding the color in a SID of the SID list present in the SRv6 header. The first solution is simple but has some drawbacks: the number of bits available in the DS field is limited (6 or 8) and they are considered precious. Using two colors we need a bit, in addition we can use a second bit to differentiate between colored traffic to be monitored and uncolored traffic not to be monitored. This can be useful to avoid comparing the full SID List to decide if a packet is part of a flow under monitoring or not. In our current implementation described in section~\ref{sec:impl} we have used two bits of the DS field.

The solution that encodes the color in a SID, exploits the fact that according to \cite{id-srv6-network-prog} an IPv6 address representing a SID is divided in LOCATOR:FUNC:ARGS. The LOCATOR part is routable and allows to forward the packet towards the node where the SID needs to be executed. The FUNC and ARG parts are processed in the node that executes the SID. In particular, the ARG part can reserve a number of bits for the alternate marking procedures. This may allow using more than two colors. This solution however has an implementation drawback: due to the variable position of these bits, implementing an hardware processing solution is much harder and can be out of reach for current chips that need to operate at line speed. Moreover periodically changing the ARG bits in a SID of a running flow can cause an interference with the SRv6 forwarding plane (e.g. for Equal Cost MultiPath) when the SID is used as IPv6 destination addresses.

\subsection{Data Collection}

The data collection in operated using the TWAMP Light protocol extension specified in \cite{gandhi-spring-twamp-srpm-08}. The Sender prepares a query message that that is sent in the congruent path with traffic, and the Reflector replies with a response message that can be sent in-band, in the reverse path of the data traffic or out-band.

TWAMP messages are inserted into an UDP/IPv6 packet that are sent with Source and Destination UDP ports configured by the user. However the the UDP destination port is used to identify the message type and the authentication mode, and since TWAMP does not have any indication to distinguish between query and response messages, source port need to be different form destination port.

The UDP packet is then encapsulated in another IPv6 packet comprising the SRv6 header as for normal data traffic.  To distinguish such packets from the normal traffic, a special policy is applied using the $END$ function $END.OP$ as defined in \cite{ali-6man-spring-srv6-oam-03} that modify the target SID to punt the packets in the egress node. In the following we provide a brief description of the simple authenticated version of TWAMP messages with the relative fields to better clarify the data collection procedure.

\begin{figure}[t!]
    \centering
    \includegraphics[width=2.4in]{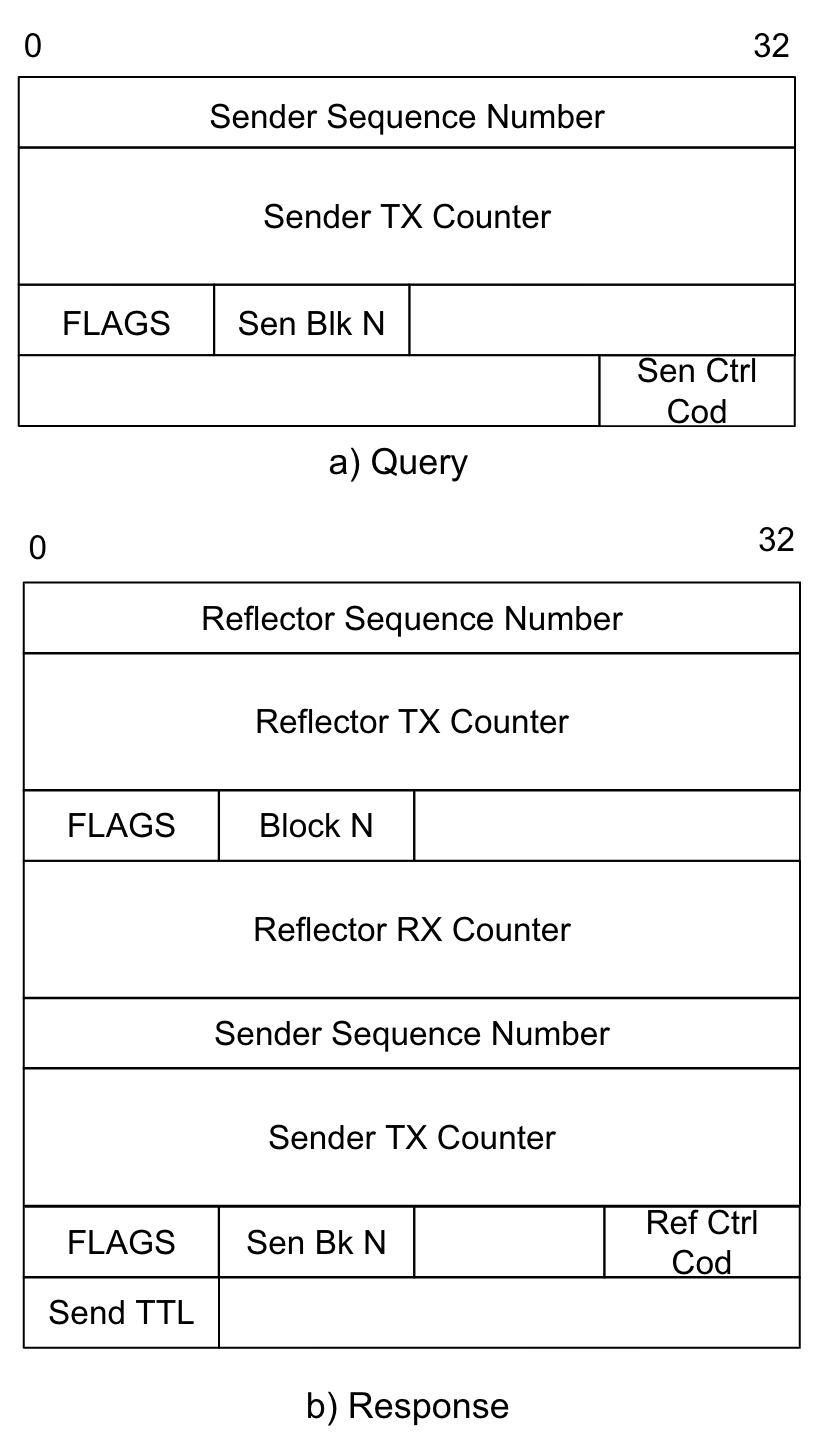}
    \caption{TWAMP Loss Measurement a) Query  and b) Response messages defined in \cite{gandhi-spring-twamp-srpm-08}}
    \label{fig:twamp-proto}
\end{figure}

\subsubsection{TWAMP Query message}

In figure \ref{fig:twamp-proto} a) is depicted the query message that includes a Sender Sequence Number (32-bit), the Sender Transmit Counter (64-bit) and the Block Number (8-bit) for the specific flow that we are monitoring. The Block Number specifies the color that we are reading and should refer to the inactive color. Flags are used to specify some options such as the counter format and the type counting mode (bytes or packets). Finally the control code (Ctrl Cod) indicates to the Reflector if the response is required in-band or out-band.

\subsubsection{TWAMP Response message}

The messages is used to collect the counters of a given color, in order to evaluate the loss. If path monitoring is bidirectional, the Reflector sends to the Sender a response packet that goes through the network in the reverse direction, collecting the counters of the return path. 

\subsubsection{TWAMP Response message}

The response message includes data collected in the Reflector and data received from the sender, i.e. the Sender Sequence Number, the Transmission Counter and Block Number.
The Reflector includes the Reception counter for the flow specified in the SR Header and if the response is in-band, the Sequence Number, the Transmission Counter and Block Number for the return path.

When the sender receives the response, it is able to calculate the loss in the forward path using the Sender TX Counter an Reflector RX Counter. It can also calculate the loss in the return path using the Reflector TX Counter and reading locally the corresponding RX counter.

\section{Monitoring System Implementation}\label{sec:impl}

The Performance Monitoring (PM) system has been implemented using several open source platforms and software frameworks. In particular an exemplary data plane has been implemented extending the Linux kernel SRv6 networking and using other Linux frameworks for packet processing, namely IPset and eBPF. 
We integrated in the system the counting solution based on IPset presented in \cite{loreti2020implementation} and a brand new implementation that exploit the eBPF Virtual Machines to execute both the packet coloring and counters. All the developed software components are available as open source \cite{netgroup-srv6-pm}.  The local SRv6 Manager defined in \cite{ventre2018sdn} is extended for controlling the TWAMP Sender and Reflector that implement the TWAMP protocol for SRv6. The SRv6 Manager provide the Southbound loss monitoring interface that allows the SDN Controller to communicate with the nodes configure, start and stop the measure. 

In the following subsections we are going to provide brief overviews on the systems and tools that we have extended and a detailed descriptions of our contributions. 

\subsection{Linux SRv6 subsystem}

The Linux kernel SRv6 subsystem \cite{lebrun2017implementing} supports the basic SRv6 operation described in \cite{ietf-6man-segment-routing-header} and most of the operations defined in \cite{id-sr-service-programming}. A Linux node can classify incoming packets and apply SRv6 policies, e.g. encapsulate the packets with an outer packet carrying the list of SRv6 segments (SIDs). A Linux node can associate a SID to one of the supported operations, so that the operation will be executed on the received packets that have such SID as IPv6 Destination Address. More details on the Linux SRv6 implementation with a list of the currently supported operations can be found in \cite{ahmedperformancefull}.

\subsection{IPset based counter} \label{ipset-framework}

IPset \cite{ipset} is an extension to Netfilter/Xtables/Iptables that is available through Xtables-addons \cite{xtables-addons}. IPset allows to create rules that match entire \textit{sets} of elements at once. Unlike normal Iptables chains, which are stored and traversed linearly, elements in the sets are stored in indexed data structures, making lookups very efficient, even when dealing with large sets.
Depending on the type, an IPset may store IP host addresses, IP network addresses, TCP/UDP port numbers, MAC addresses, interface names or combination of them in a way, which ensures lightning speed when matching an entry against a set.

As a result, a single Iptables command is required regardless of the number of elements in the set. If we want to achieve the same result with the use of Iptables only, it would be necessary to create a chain and insert as many rules as the elements contained in the set.
IPset provides different types of data structures to store the elements (addresses, networks, etc). Each set type has its own rules for the type, range and distribution of values it can contain. Different set types also use different types of indexes and are optimized for different scenarios. The best/most efficient set type depends on the situation. 
The hash sets offer excellent performance in terms of speed of the execution time of lookup/match operation and they fit perfectly to our needs.
For example inserting $N$ IP addresses in the hash set, the cost of searching for an address is asymptotically equal to $O(1)$. Conversely, the same operation with Iptables would have a cost of $O(N)$.


The actual implementation of IPset does not allow to store elements of \texttt{SID list} type within a hash set. Therefore, in order to use IPset for our purposes we have patched it by creating a new hash set called \texttt{sr6hash} so that we can insert the SID lists that we want to monitor. To support the new \texttt{sr6hash} hash type, we patched the IPset framework on both the user-space and the kernel-space sides. In the user-space side, we defined a new data structure, the \texttt{nf\_srh}, which contains the SRH header with a SID list whose maximum length is fixed and set at compilation time (16 SIDs in our experiments). Details on the counter implementation and its performance results are reported in \cite{loreti2020implementation}.

\subsection{The extended Berkeley Packet Filter (eBPF)}

Proposed in the early ‘90s, the  Berkeley Packet Filter (BPF) \cite{bsd-packet-filter} is designed as a solution for performing packet filtering directly in the kernel of BSD systems. BPF comes along with its own set of RISC-based instructions used for writing packet filter applications and it provides a simple Virtual Machine (VM) which allows BPF filter programs to be executed in the data path of the networking stack. The bytecode of an BPF application is transferred from the userspace to the kernel space where it is checked for assuring security and preventing kernel crashes. BPF also define a packet-based memory model, few registers and a temporary auxiliary memory.

The Linux kernel has supported BPF since version 2.5 and the major changes over the years have been focused on implementing dynamic translation \cite{jit-packet-filters} between BPF and x86 instructions. Starting from release 3.18 of the Linux kernel, the Extended BPF (eBPF) \cite{linux-socket-filtering} represents an enhancement over BPF which adds more resources such as new registers, enhanced load/store instructions and it improves both the processing and the memory models. eBPF programs can be written using assembly instructions later converted in bytecode or in restricted C and compiled using the LLVM Clang compiler. The bytecode can be loaded into the system through the \texttt{bpf()} syscall that forces the program to pass a set of sanity/safety-checks in order to verify if the program can be harmful for the system. Only safe eBPF programs can be loaded into the system, the ones considered unsafe are rejected.
To be considered safe, a program must meet a number of constraints such as limited number of instructions, limited use of backward jumps, limited use of the stack and etc. These limitations \cite{complex-network-service-with-ebpf} can impact on the ability to create powerful network programs.

eBPF programs are triggered by some internal and/or external events which span from the execution of a specific syscall up to the incoming of a network packet. Therefore, eBPF programs are hooked to different type of events and each one comes with a specific execution context. Depending on the context, there are programs that can legitimately perform socket filtering operations while other can perform only traffic classification at the TC layer and so on.
eBPF programs are designed to be stateless so that every run is independent from the others.
The eBPF infrastructure provides specific data structures, called \textit{maps}, that can be accessed by the eBPF programs and by the userspace when they need to share some information.

\subsection{The eBPF based counters implementation}

The Loss Monitoring implementations presented in \cite{loreti2020implementation} were based on some Linux kernel modules which carried out all the needed operations for parsing packets, updating flow counters and coloring. Kernel modules allowed us to implement the counting system in a very effective way, but this approach comes with a number of disadvantages that we briefly report hereafter. Every time we need to update the Linux kernel, we also have to maintain the counting modules updated. Furthermore, if we want to add a functionality to our counting implementation, we have to remove the loaded modules and to replace them with the updated ones.
Replacing a module is not an atomic operation: the time between the removal and the loading of the updated module causes a down-time of the monitoring service which may not be tolerated if we are accounting statistics on high-rate data flows. A kernel module can have unconditional access to most of the internals of the system and, if there is a bug, this can compromise the overall system security and stability. In addition, the Iptables and IPset modules are hooked to the Netfilter which can not guarantee the maximum performance in terms of packet processing.

An eBPF network program, despite its limitations, can solve the problems that arise by implementing the counter using kernel modules. Therefore, we designed two eBPF programs attached to specific networking hooks with the purpose of entirely replacing the previous implementations.

Both the \texttt{tc\_srv6\_pfplm\_egress}  and the \texttt{xdp\_srv6\_pfplm\_ingress} eBPF programs are written in C and compiled through the LLVM Clang compiler. The result is a single object file called \texttt{srv6\_pfplm\_kern.o} where each program is placed in a specific text section so that it can be loaded independently from the other. The two eBPF programs are designed to account individually the ingress and egress flows that match some given SR policies taking into account also the color marking.


Flows lookup and match operations are fast and efficient thanks to the \texttt{BPF\_MAP\_TYPE\_PERCPU\_HASH} map type which help us to retrieve the flow counters using the SID list as key in a lock-free fashion and in constant time, regardless of the number of policies present in the table. Currently, all the hashtable data structures provided by the eBPF infrastructure support fixed length keys. Thus, to maximize the lookup performance and minimize memory waste, we created $N$ maps/hashtables and in each map we store all the flows with the same SID list length. The value of $N$ can be decided at compile time and in our implmentation the default max SID list length is equal to $N=16$ SIDs and therefore, we come up with a total of 16 maps/hashtables. Furthermore, we also have decided to store and keep separated the ingress flows to be monitored from the egress ones, hence the total number of flow maps/hashtables is doubled.


All our maps are persistent and accessible as files through the \textit{/sys/fs/bpf/pfplm} directory and for interacting with them we provided an User API (UAPI) released in the form of a shared library. Our UAPI hides the complexity of the underneath data structures and allows the user to change the active color for marking packets, to easily add/remove flows in ingress/egress directions and to retrieve flow statistics.

The \texttt{xdp\_srv6\_pfplm\_ingress} eBPF program is hooked on eXpress Data Path (XDP) which allows intercepting RX packets right out of the NIC driver and (possibly) before the socket buffer (\texttt{skb}) is allocated.

This eBPF network program parses the headers of every incoming packet on a specific network interface, it extracts the flow (if any) and it determines, by looking at the right flow map/hashtable, whether the flow has to be monitored or not.

The \texttt{tc\_srv6\_pfplm\_egress} eBPF program leverages the hook offered by linux Traffic Control (TC) for intercepting the TX packets just before they leave the node. We attached the eBPF program to the TC hook because, unlike XDP that intercepts only RX packets, it also allows dealing with TX packets.




\subsection{TWAMP Sender and Reflector}

To test the counting system, we implemented an SRv6 version of the TWAMP Sender and Reflector using the Python language implementing the procedures for counters collection. The two components integrate a driver to control EPBF or IPSet based counters and both the Sender and the Reflector periodically change the active color. 

The TWAMP messages have been implemented the using the Scapy project \cite{scapy} a python library for sending and receiving packets that also SRv6. The measurement procedure is initiated by the Sender that reads the local counter and generate the query packet that is sent using the SRv6 path. The Reflector receives the packet, reads the missing counters and sends the response packet back to the Sender, that eventually is able to evaluate the packet loss. The open-source implementation of the python code is available online \cite{netgroup-srv6-pm}. 

The TWAMP Sender and Reflector are interfaced by the SDN controller through the gRPC interface described in the following section.

\subsection{Southbound gRPC APIs for Loss Monitoring}
The SRv6 nodes can be controlled by the Southbound gRPC APIs, that can be logically divided in two gRPC services: i) SRv6Manager, that provides functionality for the management of SRv6 entities like path and behaviors; and ii) SRv6PM that provides functionality for performance measurement. 

SRv6Manager is implemented through a set of APIs which provide CRUD operations for SRv6 entities.
Further information on SRv6Manager is available in \cite{ventre2018sdn}, while the full set of implemented APIs is documented in appendix \ref{a:rpc_manager}. SRv6PM service implements a basic set of operation required for monitoring traffic flows. This set includes function calls for:
\subsubsection{Start monitoring a flow} both the Sender and the Reflector are activated by the controller when a flow need to be monitored; the call parameters configure the measurement, providing information such as the monitored SID Lists and the Measure ID, or carrying the basic TWAMP configuration, such as UDP ports.
\subsubsection{Stop monitoring a flow} both the Sender and the Reflector are instructed to stop the data communication, as is no longer necessary to monitor a specific flow.
\subsubsection{Retrieve Measured Data} the Sender is put in receiving mode, so to transfer on demand information of the recent measurements collected by the TWAMP sessions.

Appendix \ref{a:rpc_pm} presents further details on the provided methods and on the implemented gRPC protocol.

\section{SRv6-PM open source platform and testbeds} \label{sec:testbeds}

In this section we first presents the open source PM platform and the cloud-native approach used to implement the SDN control and management planes. Then we will describe the three testbed facilities used to validate the proposed architecture and to evaluate the performance of our implementations.

\begin{figure*}[t!]
    \centering
    \includegraphics[width=0.9\linewidth]{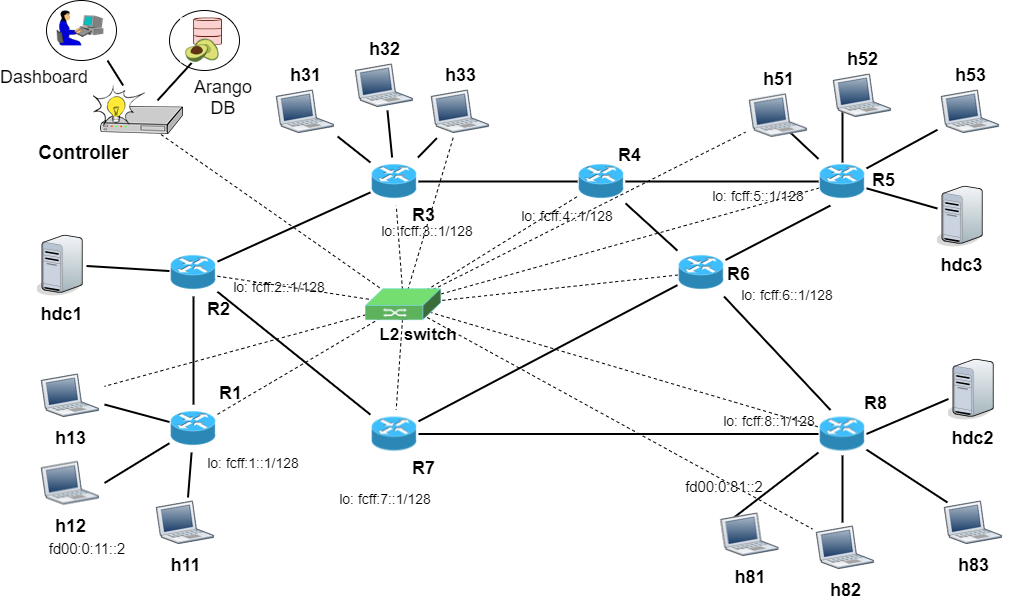}
    \vspace{-0.9cm}
    \caption{Mininet testbed topology}
    \label{fig:testbedtop}
    \vspace{-0.5cm}
\end{figure*}

\subsection{Reproducing the experiments and CI/CD}\label{sec:cicd}

In this section we describe the characteristics of the released open source platform, called SRv6-PM, that allows replicating the experiments and developing new features on top of the existing ones. Our aim is not only to release a \dq{working copy} of our environment to reproduce our results. Rather, we offer an extendable platform that can be used to develop new features, and can be extended and adapted to support additional scenarios. The platform is based on the Continuous Integration / Continuous Development (CI/CD) principles typical of a cloud-native approach. 

The SRv6-PM platform is made up of two parts: the first part includes the SRv6-PM data plane, the second part the SDN control and management planes, including the stack of components described in sec.~\ref{sec:archsol}. As for the SRv6-PM data plane, we emulate a number of SRv6 Linux based routers using the mininet \cite{mininet} network emulator and we provide a ready-to-go Virtual Machine, as discussed further in sec.~\ref{sec:mininet-testbed}.

In this section, we discuss how we deploy our cloud-native workload implementing the SDN control and management planes of the SRv6-PM platform. We start with a Dockerfile defining the containerized version of each basic component. We have a \dq{manual} and \dq{static} approach for the deployment, which would be enough for the plain replication of the whole platform. In this approach, multiple components are aggregated in a deployment unit to be run with the Docker Compose tool (described with a docker-compose.yml file). The Docker Swarm tool is used to deploy the various deployment units over the hosts running Docker. In the simplest case, all deployments units can be deployed on the same host, otherwise each deployment unit can be run on a different host. Of course the proper configuration of the networking is needed to allow the different deployment units to communicate each other, and the deployment unit that contains the controller to communicate with the SRv6-PM data plane.

This manual and static approach is not the preferred way to deploy a cloud-native workload, as it is inflexible and might require a lot of effort for any reconfiguration or extensions (e.g. for adding new components in the stack). On the contrary, a typical cloud-native deployment is based on a container orchestration platform, such as Kubernetes \cite{kubernetes}. For these reasons, we offer a second approach to orchestrate the workloads that compose our demonstrator and experiments, which is \dq{automatic} and \dq{dynamic}.

In particular, we extended the \textit{Kubernetes Playground} platform \cite{kubernetes-playground} to deploy the cloud-native part of SRv6-PM. Kubernetes Playground is an automated container deployment solution developed on top of Kubernetes \cite{kubernetes}, which is capable of:

\begin{enumerate}
\item Provisioning and configuration of a virtualized runtime environment (i.e. it includes a full installation of kubernetes from scratch).
\item Orchestrating workloads in the virtualized runtime environment.
\end{enumerate}

Kubernetes Playground uses Vagrant \cite{vagrant} to provision the runtime environment, and supports different hypervisors, such as Oracle VirtualBox \cite{oracle-virtualbox}, KVM \cite{kvm}, and Hyper-V \cite{hyperv}. With the vagrant-libvirt \cite{vagrant-libvirt} \cite{vagrant} plugin, we were able to abstract away the implementation details of the supported hypervisors by leveraging libvirt \cite{libvirt} programming interfaces, instead of having to deal with each hypervisor separately. Additionally, we extended Vagrant and vagrant-libvirt to support multiple physical libvirt hosts for the same runtime environment, provided that the libvirt hosts can communicate with each other. This feature let administrators benefit of greater deployment flexibility. For example, they might deploy part of the VMs of a cluster in a libvirt host, while deploying the remaining VMs on a second libvirt host.

Since the early stages of design and development of this platform, we applied a test driven development methodology, so that we could automatically test, verify, and validate every single component of the solution, including the underlying virtualized infrastructure, and the solution as a whole. Every single change to the code base is reviewed by multiple core developers, and is checked against a set of style guides, best practices, and test suites. In summary, we developed an automated CI/CD pipeline for Kubernetes Playground that covers the platform, the virtualized infrastructure that supports the platform, and the workloads that we deploy on the platform.

Kubernetes Playground offers a configuration mechanism that supports the customization of the runtime environment. For example, administrators can choose how many nodes are part of a Kubernetes cluster, and which Kubernetes CNI plugin \cite{kubernetes-cni} should be deployed in the cluster.

Note that the use of Kubernetes Playground is not mandatory for the setup of an SRv6-PM platform instance and the replication of our experiment. The \dq{dockerized} version of the stack is available online in the project repository \cite{netgroup-srv6-pm}. It is always possible to use the manual approach of deploying containers using the basic Docker tools, or to import the containers in an existing Kubernetes deployment (starting from the provided Dockerfiles and Kubernetes descriptors).

\subsection{Mininet Testbed}
\label{sec:mininet-testbed}

To validate the proposed SRv6-PM solution we devised a complete proof of concept using the mininet \cite{mininet} testbed of the ROSE project  \cite{rose}. ROSE (Research on Open SRv6 Ecosystem) offers an handy environment to run SRv6 experiments through a virtual machine (called rose-srv6 VM) \cite{rose-vm}. The ROSE mininet testbed includes several ready to use configurations available inside the rose-srv6 VM.
We selected for our experiments the topology depicted in figure \ref{fig:testbedtop}, which comprises 8 SRv6 capable routers and 15 hosts (of which 12 customer hosts and 3 provider hosts). We use the open source FRR routing suite \cite{frr}. Nodes are automatically configured through the Zebra \cite{zebra} daemon, and routing is entrusted by the IS-IS routing protocol. 
The figure shows also the IPv6 configuration of loopback routers interfaces and of two customer hosts.
The testbed topology can also include the controller as shown in \ref{fig:testbedtop}. The controller is running in a dedicated network namespace and it is located outside of the data plane emulated network. It uses an out-of-band management network (implemented with a virtual switch) to connect to all nodes, so that the communications between the controller and the nodes does not interfere with the data plane communications among the nodes. It is also possible to connect the controller to one or more virtual routers and emulate an in-band connection between the controller and the nodes. Running the controller inside the rose-srv6 VM is convenient for running simple networks scenarios. In the more general case the rose-srv6 VM is used for a full  emulation of the dataplane, while the controller and the other monitoring/ big-data tools are run in a separate execution infrastructure (e.g. a cloud infrastructure, or a private Kubernetes cluster). 


Inside the ROSE VM it is possible to setup bidirectional SRv6 tunnels among the customer hosts to enable their communication, both manually using the Linux CLI, or using our controller. For example the controller can add an SRv6 policy that connects routers R1 and R8 specifying R3 and R4 as "waypoints". Thus the packets that must reach network 8 from network 1, will be encapsulated in an SRv6 flow that will pass through nodes 3 and 4

In the emulated network inside the rose-srv6 VM we are able to generate controlled traffic using the iperf3 tool and we can artificially add loss or delay  using Linux \texttt{tc} and \texttt{netem} tools.


\subsection{Integration with Cisco Testbed}
\begin{figure}[b!]
    \centering
    \includegraphics[width=0.48\textwidth]{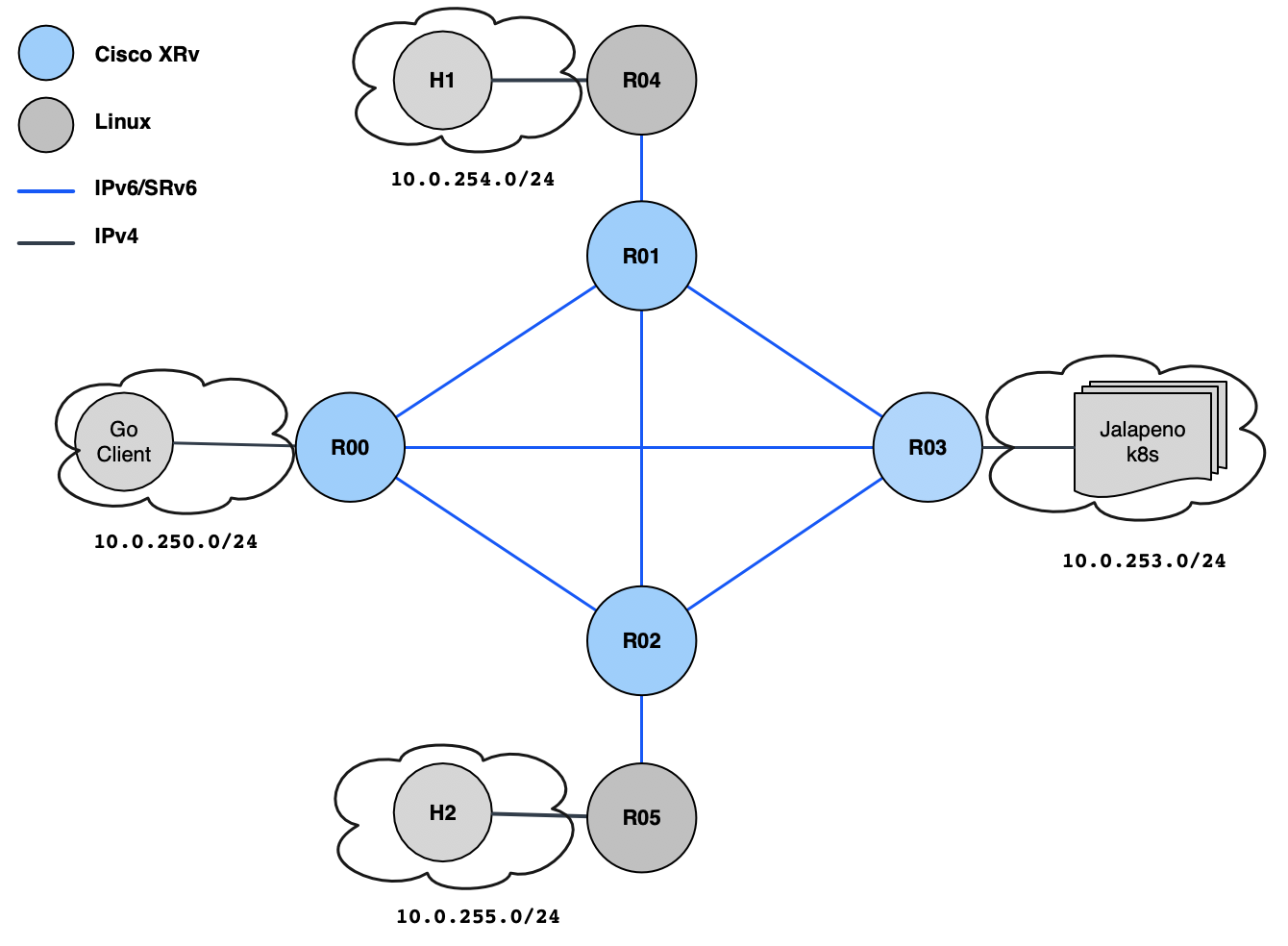}
    \caption{Cisco testbed topology}
    \label{fig:cisco_testbed}
    \vspace{-3ex}
\end{figure}

In order to show the portability and modularity of our performance monitoring framework to other SRv6 networking facilities, we integrated several components of the architecture shown in Figure~\ref{fig:bigdataarch} in the testbed depicted in Figure~\ref{fig:cisco_testbed} and deployed in the  Cisco lab facilities. It is composed of four Cisco virtual routers (R00-R03) running Cisco IOS XR routers as highlighted by the blue nodes. These routers are running the IS-IS protocol as routing protocol, and the BGP protocol to advertise SRv6 services such as VPN. The four routers (R00-R03) are configured to export their telemetry data and BGP topology information, using the BGP Monitoring Protocol (BMP)~\cite{rfc7854}, to the the Jalapeno framework~\cite{jalapeno}. 

Jalapeno is a cloud-native SDN infrastructure developed by Cisco. It comes as a Kubernetes package and is deployable on bare metal, VM, and on Google Cloud. In this testbed, we deployed Jalapeno using MicroK8s~\cite{microk8s} on a baremetal server. It includes several microservices including: (1) collectors which capture network topology and performance data and feed the data to Kafka; (2) Data Processors, Graph Database (ArangoDB), and Time-Series Database (InfluxDB); (3) Grafana for data visualization.

To integrate the SRv6-PM monitoring in the Cisco testbed, we introduced two Linux routers (R04 and R05) which we use for the packet loss measurements. The Linux routers leverage the implementation described in Section~\ref{sec:impl} and are managed through the SRv6 Performance monitoring control plane available at~\cite{netgroup-srv6-pm}. The Linux routers routers (R04 and R05) routes are integrated with the core routers (R00-R03) at the IP routing level using the IS-IS dynamic routing protocol. In fact, R04 and R05 run an open-source routing daemon (FRRouting~\cite{frr} allowing them to advertise their IPv6 prefixes using the IS-IS protocol. 

We leverage the InfluxDB of the Jalapeno framework to store the packet loss measurements between R04 and R05. We deployed the dockerized SRv6-PM Controller inside the Jalapeno Kubernetes infrastructure, The SRv6-PM collects the packet loss measurements from R04 and R05 and writes them to Kafka. Our Telegraf processor (that we ported as well to Jalapeno) reads the measurements from Kafka and writes them into the InfluxDB to make them available for any application using the Jalapeno infrastructure. 

The integration of the SRv6-PM solution into the CISCO testbed was fully successful thanks to the cloud-native architecture of our SRv6-PM solution. The integration process comprises 1) deploying the Linux routers that do the packet loss measurements in the CISCO testbed; 2) adapting the Kubernetes descriptors from the SRv6-PM reference platform to jalapeno infrastructure; 3) deploying the containerized applications as part of the Jalapeno Kubernetes infrastructure. The step 1) took few hours (note that this step is not based on cloud-native technologies), the step 2) less than 1 hour, the step 3) just few seconds. 

\subsection{Cloudlab Testbed for processing load evaluation}
\label{sec:cloudlab-testbed}

In order to evaluate the impact of the proposed solutions in a real Linux implementation described in the implementation section, we set up a testbed according to RFC 2544 \cite{rfc2544}, which provides the guidelines for benchmarking networking devices.

\begin{figure}[t!]
    \centering
    \includegraphics[width=0.48\textwidth]{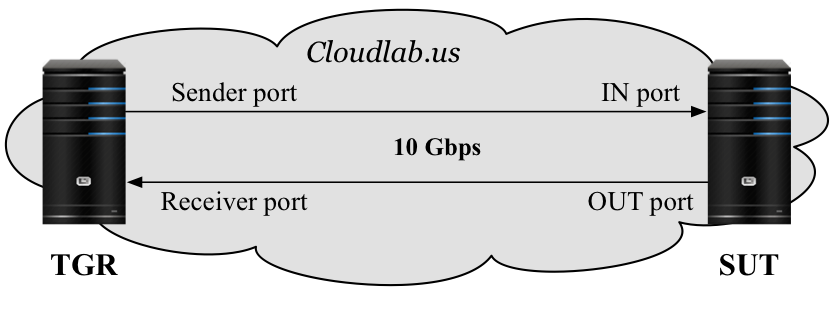}
    \caption{Testbed architecture}
    \label{fig:testbed}
    \vspace{-3ex}
\end{figure}

Figure~\ref{fig:testbed} depicts the used testbed architecture, made of two nodes denoted as \textit{Traffic Generator and Receiver (TGR)} and \textit{System Under Test (SUT)}.  
In our experiments we only consider the traffic in one direction: the packets are generated by the TGR on the Sender port, enter the SUT from the IN port, exit the SUT from the OUT port and then they are received back by the TGR on the Receiver port. Thus, the \textit{TGR} can evaluate all different kinds of statistics on the transmitted traffic including packet loss, delay, etc. 

The testbed is deployed on the CloudLab facilities \cite{ricci2014introducing}, a flexible infrastructure dedicated to scientific research on the future of Cloud Computing. Both the TGR and the SUT are bare metal servers with Intel Xeon E5-2630 v3 processor with 16 cores (hyper-threaded) clocked at 2.40GHz, 128 GB of RAM and two Intel 82599ES 10-Gigabit network interface cards. The SUT node runs a vanilla version of Linux kernel 5.4 and hosts the PF-PLM system. We consider two configurations to evaluate the SUT performance:
\begin{enumerate}
\item the SUT is configured as the ingress node of the SRv6 network, i.e. it executes the encapsulation operation.
\item the SUT is configured as the egress node of the SRv6 network, i.e. it executes the decapsulation operation.
\end{enumerate}

In the TGR node we exploit TRex \cite{trex-cisco} that is an open source traffic generator powered by DPDK \cite{dpdk}. 
We used SRPerf \cite{ahmedperformance}, a performance evaluation framework for software and hardware implementations of SRv6, which automatically controls the TRex generator in order to evaluate the maximum throughput that can be processed by the SUT. The maximum throughput is defined as the maximum packet rate at which the packet drop ratio is smaller then or equal to 0.5\%. This is also referred to as Partial Drop Rate at a 0.5\% drop ratio (in short PDR@0.5\%). 
Further details on PDR and insights about nodes configurations for the correct execution of the experiments can be found in \cite{ahmedperformance}.

\section{Functional validation and Performance Results}

\subsection{Functional validation of Loss Monitoring}

Using the mininet testbed described in sec.~\ref{sec:mininet-testbed}, we performed several experiment to validate the accurate loss monitoring system. Thus we designed an experiment with 12 simultaneous flows among the hosts in the four different networks 1,3,5 and 8.
The controllers create policies for the 12 flows according to the table \ref{tab:my-table}. For example if we consider the path between R1 and R8, the controller set two way points: R2 and R7. As can be noted, all the defined paths are symmetric, i.e. the forward and the reverse paths have the same set of waypoints but in the reverse order. 

\begin{table}[h!]
\centering
\caption{Waypoints for the different paths of the SRv6 policies}
\label{tab:my-table}
\begin{tabular}{c|c|c|c|c|}
\cline{2-5}
\textit{\textbf{}}                & \textit{\textbf{R1}} & \textit{\textbf{R3}} & \textit{\textbf{R5}} & \textit{\textbf{R8}} \\ \hline
\multicolumn{1}{|c|}{\textbf{R1}} & -                    & R2                   & R3, R4               & R2, R7               \\ \hline
\multicolumn{1}{|c|}{\textbf{R3}} & R2                   & -                    & R4, R6               & R6                   \\ \hline
\multicolumn{1}{|c|}{\textbf{R5}} & R3, R4               & R4, R6               & -                    & R6                   \\ \hline
\multicolumn{1}{|c|}{\textbf{R8}} & R2, R7               & R6                   & R6                   & -                    \\ \hline
\end{tabular}
\end{table}

To add loss we introduced with netem an artificial packet loss of $0.1\%$ in all interfaces of nodes R2 and R6. As can be noted all the paths defined in \ref{tab:my-table} goes through  R2 or R6, making all flows experience the same loss. We monitor each flows for 1 minute, that corresponds to six colors interval since we selected $T=10$.

\begin{figure}[t!]
    \centering
    \includegraphics[width=0.48\textwidth]{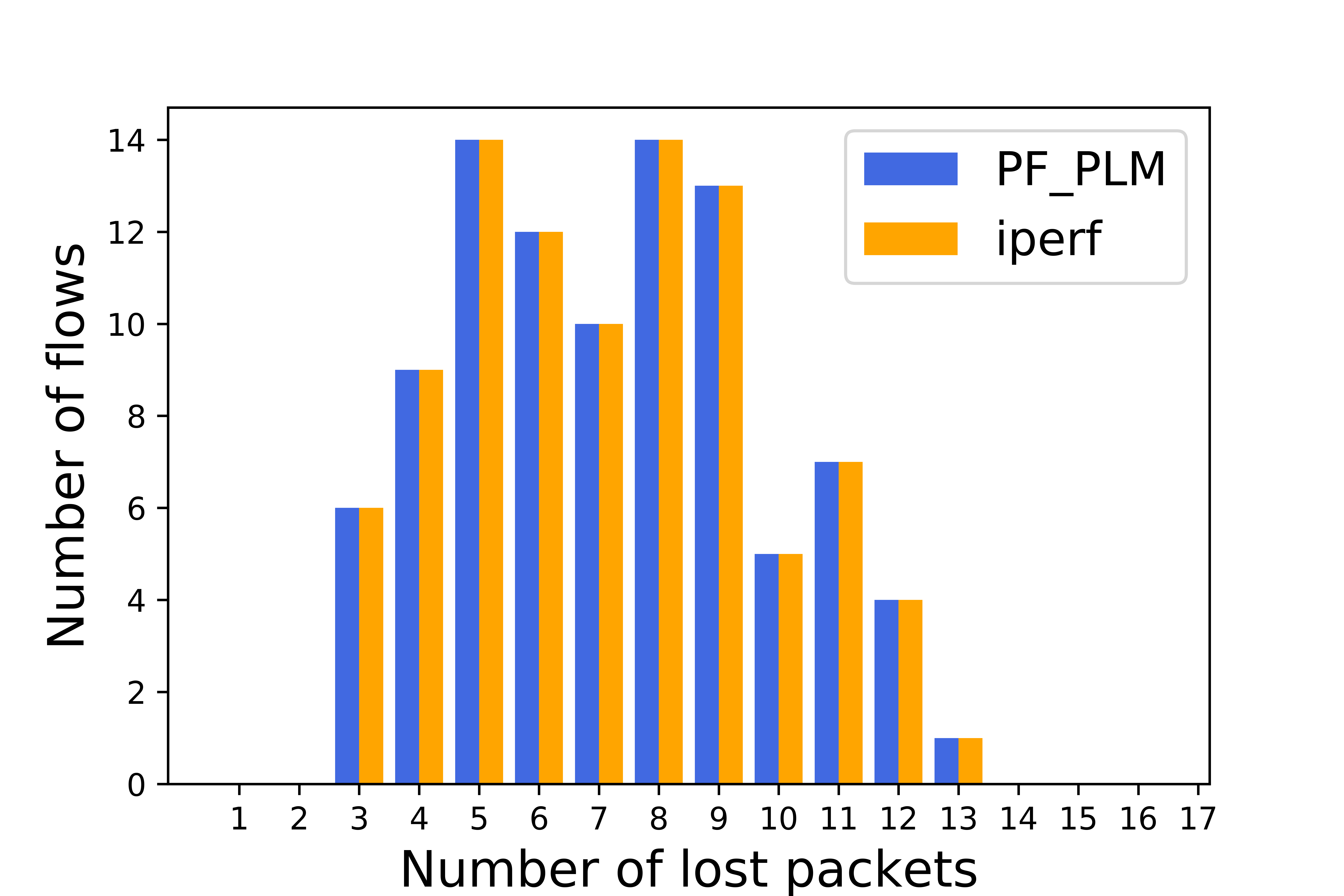}
    \caption{Histogram of measured packet loss}
    \label{fig:res1-demo-loss}
    \vspace{-3ex}
\end{figure}

In figure \ref{fig:res1-demo-loss} we plot the number of flows that experienced a certain number of packet loss comparing the results obtained by iperf and PF-PLM tools respectively. Clearly, both tools measure the same loss and this shows that the counting system is as accurate as the traffic generator.

\subsection{Processing load of packet counting}
\label{sec:sr_test}
\label{sec:perf}

Considering the testbed described in sec.~\ref{sec:cloudlab-testbed}, we carried out the experiments to evaluate the processing load of packet counting for loss monitoring. We crafted IPv6 UDP packets encapsulated in outer IPv6 packets (78 bytes including all headers). The outer packets have an SRH with a SID list of one SID. We repeated each test four times (note that, as described in  \cite{ahmedperformance}, each test includes a large number of experiments and repetitions to estimate the maximum throughput).

\begin{figure}[b!]
    \centering
    \includegraphics[width=0.48\textwidth]{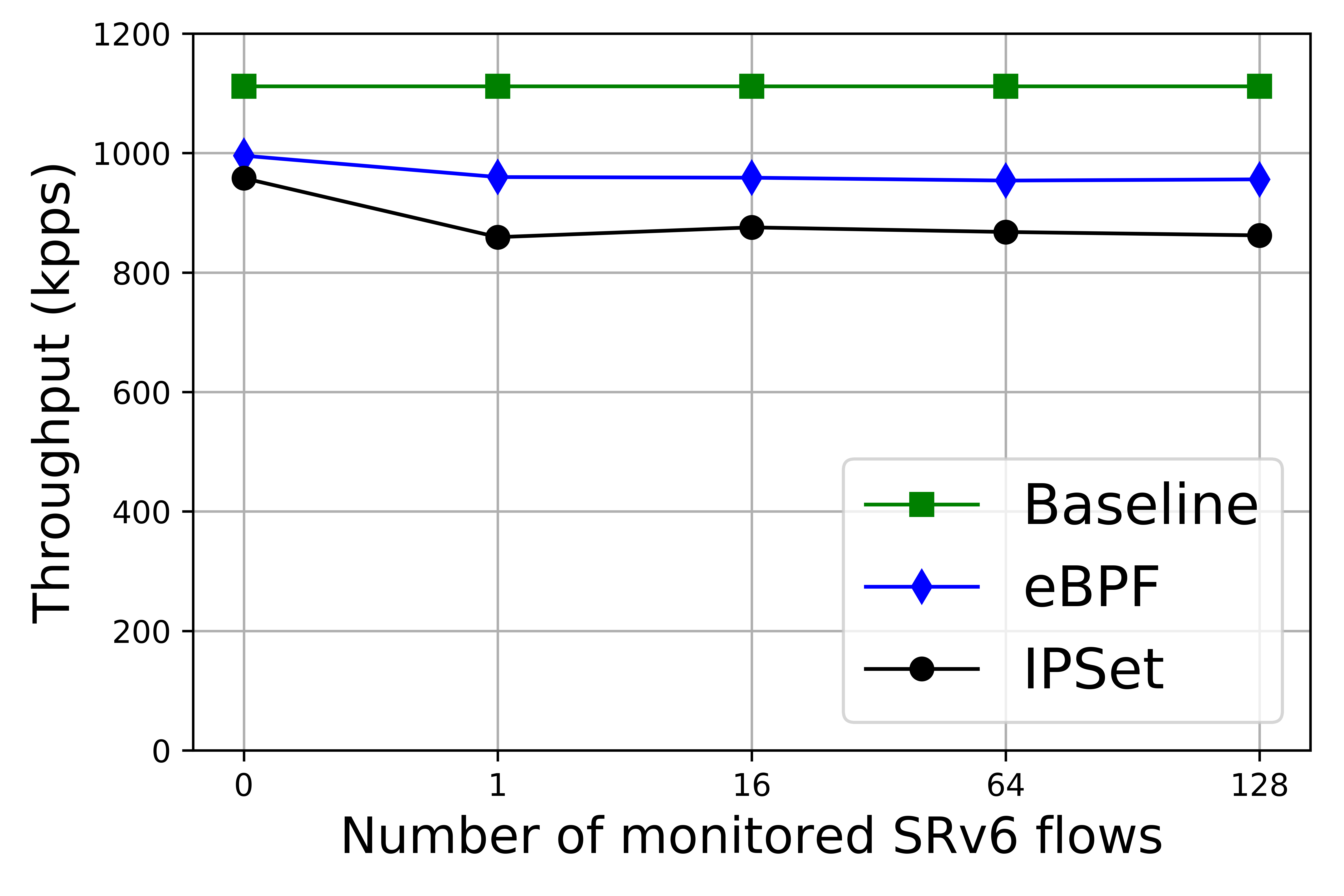}
    \caption{SUT throughput (ingress node configuration)}
    \label{fig:res1-decap}
    \vspace{-3ex}
\end{figure}

On the ingress node, the eBPF program hooked to TC allows to achieve better performance in terms of the overall throughput compared to the performance obtained with Netfilter / IPset.
The eBPF based PF-PLM starts with a 10.5\% degradation (w.r.t. the baseline) when there are zero flows to be monitored. The overall performance degradation reaches the average value of 13.9\% for one flow to be monitored and this value remains stable regardless of how many additional flows are added.
The IPset based PF-PLM starts with a 13.8\% performance degradation for zero flows to be monitored and it reaches the stable value of 22.0\% for one or more monitored flows.
This is due to the low processing overhead of the TC layer compared to the non-negligible one introduced by the Netfilter layer on which are hooked many running subsystems, i.e: iptables and all the related tables and chains.

On the egress node, the performance achieved with the eBPF program are considerably better if compared to performance of the IPSet module.
The eBPF based PF-PLM achieves a 4.2\% of performance degradation (w.r.t. the baseline) in case there are no flows to be monitored. The performance drops at 5.4\% on average if there is one flow to be monitored and it keeps stable no matter of how many flows (to be monitored) are subsequently added.
The IPset based PF-PLM introduces a 15.0\% of degradation for zero flows and it remains stable at 17.6\% for one or more flows to be monitored.
The reason lies in the fact that the eBPF program is directly hooked to the eXpress Data Path (XDP) and, in our case, the network card driver supports the XDP native-mode. Native mode executes eBPF/XDP program in the networking driver early RX path. At this stage, a network packet is nothing more than a mere sequence of bytes with no associated metadata. Raw data access is very efficient, as the operations needed for parsing the packet headers and for extracting the flows are not so computationally expensive.

\begin{figure}[t!]
    \centering
    \includegraphics[width=0.48\textwidth]{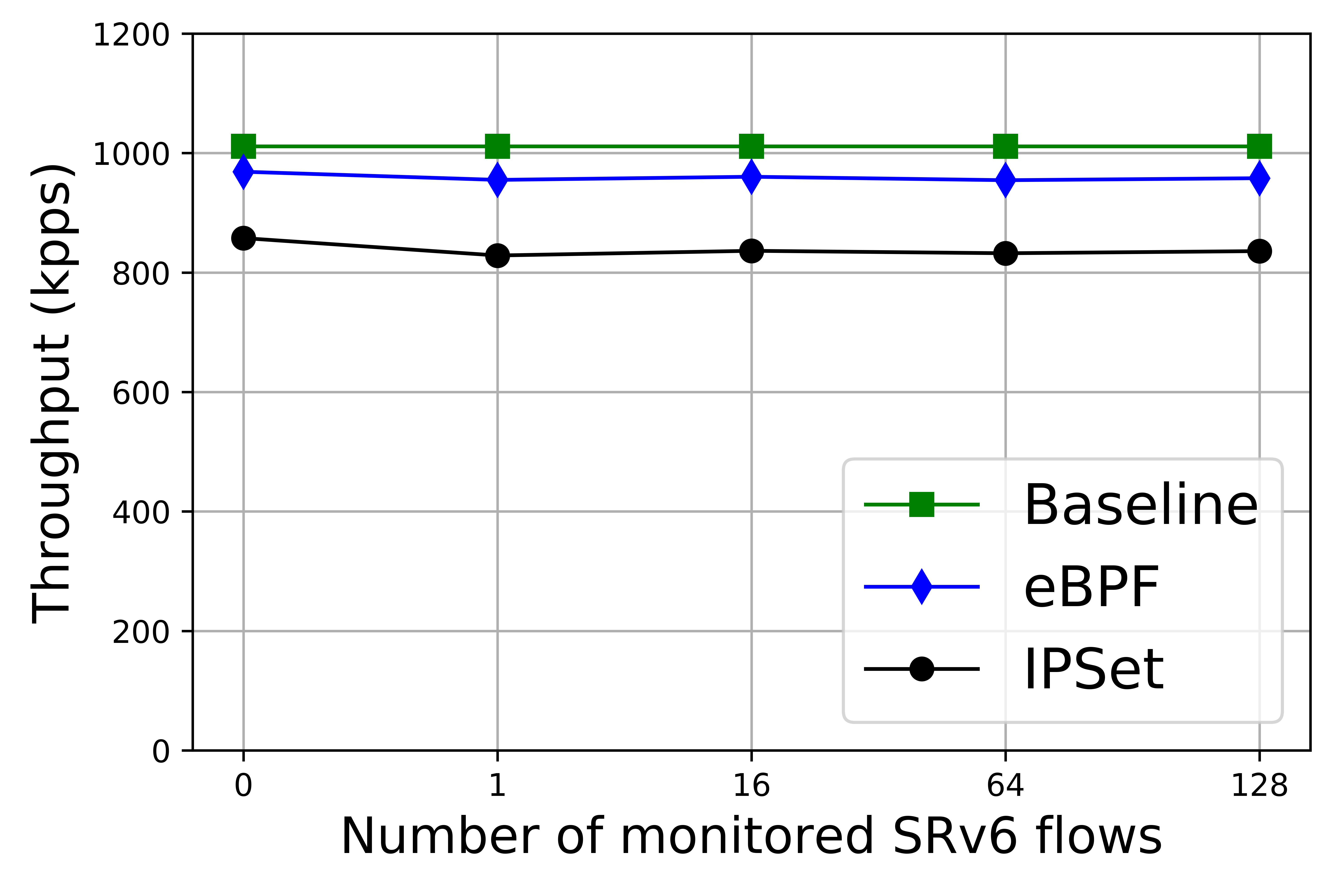}
    \caption{SUT throughput (egress node configuration)}
    \label{fig:res3-decap}
    \vspace{-3ex}
\end{figure}

\section{Conclusions}

In this paper we have first analyzed the status of the ongoing efforts for the standardization of Performance Monitoring (PM) in SRv6 networks, focusing on loss monitoring. We have selected a proposal for the data plane measurement protocol in SRv6 based on the extension of TWAMP protocol and implemented it in Linux. Then we have designed the control interactions between the SDN controllers and the nodes to setup the Per-Flow loss measurements on a set of SRv6 flows and to collect the measurement results. Accordingly, we defined and implemented the gRPC API between the SDN controller and the nodes. Our next step has been to design the cloud native architecture for the aggregation, processing and storage of monitoring data. This is based on open source components for Big Data processing (Kafka, Telegraph, Influx DB, Grafana, ArangoDB). Our SDN controller is also part of this cloud-native stack of components. 

On the Data plane, the proposed solutions for Loss Monitoring consider the \textit{alternate marking} method to achieve an accurate evaluation of packet loss (single packet loss granularity). We started from these proposals and we have defined a packet marking mechanism based on the DSCP Byte (Diffserv Code Point). We have introduced our eBPF based implementation of Per-Flow Packet Loss Monitoring (PF-PLM) in Linux. We compared this solution in terms of processing load with respect to the best implementation based on ipset that we had presented in our previous paper. We were able to achieve a significant increase of performance with respect to the ipset implementation, getting closer to the baseline maximum throughput obtained for a flow which is not monitored.

We demonstrated the correct behavior of the SRv6-PM data plane implementation with extensive experiments in which we monitored a number of flows while applying a synthetic packet loss ratio. The loss measurement data collected by our SRv6-PM monitoring solution perfectly matches the loss measured at the application level by traffic generator tool (iperf).

We believe that the SRv6-PM open source cloud-native platform represents a valuable re-usable tool for the further development and improvements of Performance Monitoring in SRv6 networks. This platform can be used to assist in the process of standard definition, by allowing the early prototyping and testing. For example we are planning the development of SRv6 Delay Monitoring, which also has been proposed in the standardization.

\section*{Acknowledgments}
This work has received funding from the Cisco University Research Program Fund and from the EU H2020 5G-EVE project.

\bibliographystyle{IEEEtran}
\bibliography{biblio.bib}
\clearpage

\appendices
\section{Southbound gRPC API}
\label{a:rpc_manager} \label{a:rpc_pm}

\begin{figure*}[htbp]
    \centering
    \includegraphics[width=0.78\textwidth]{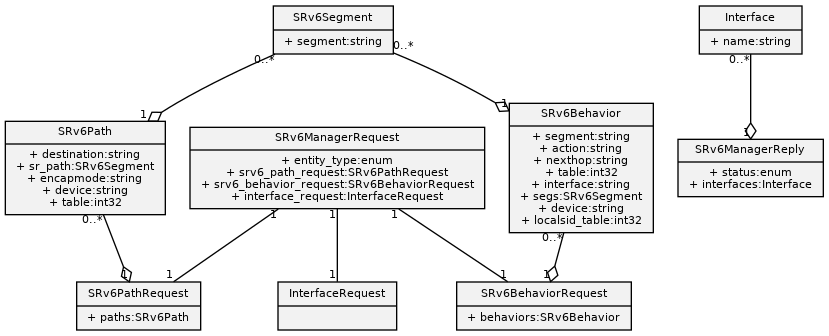}
    \caption{SRv6Manager}
    \label{fig:srv6_manager_pb2}
    \vspace{-3ex}
\end{figure*}

\begin{figure*}[htbp]
    \centering
    \includegraphics[width=0.78\textwidth]{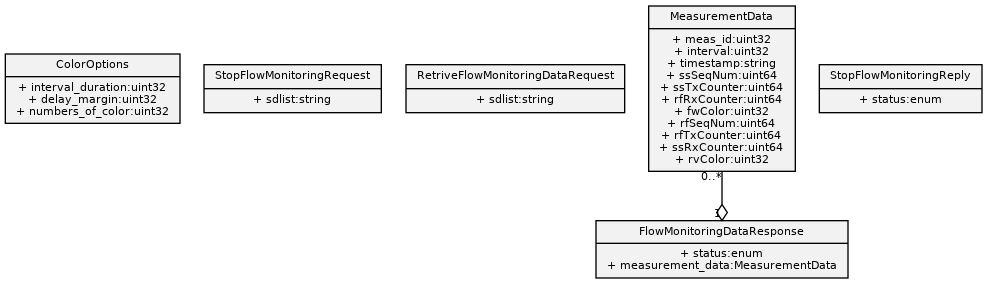}
    \caption{srv6pmCommons}
    \label{fig:srv6pmCommons_pb2}
    \vspace{-3ex}
\end{figure*}

\begin{figure*}[htbp]
    \centering
    \includegraphics[width=0.78\textwidth]{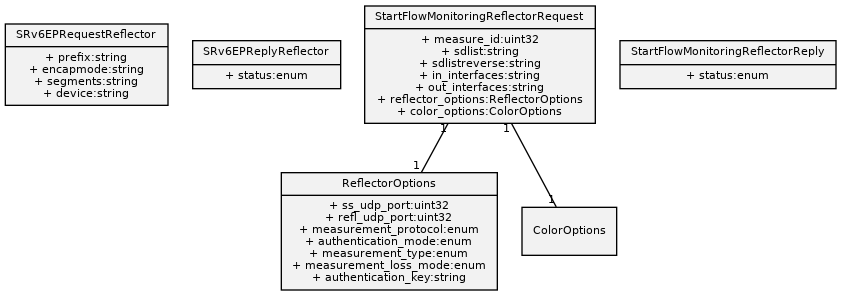}
    \caption{srv6pmReflector}
    \label{fig:srv6pmReflector_pb2}
    \vspace{-3ex}
\end{figure*}

\begin{figure*}[htbp]
    \centering
    \includegraphics[width=0.78\textwidth]{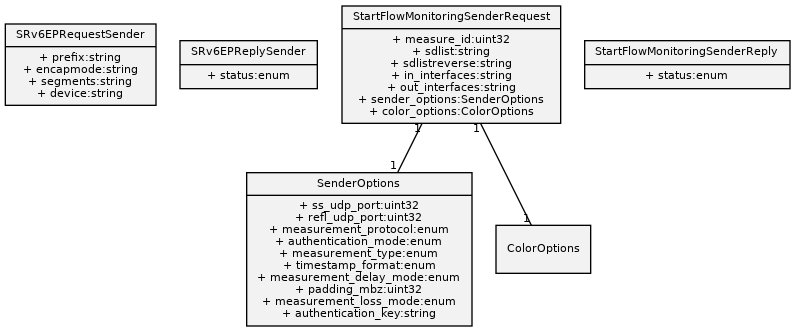}
    \caption{srv6pmSender}
    \label{fig:srv6pmSender_pb2}
    \vspace{-3ex}
\end{figure*}

In this appendix we provide the specification of the gRPC southbound API, by including the protobuf source files and the corresponding set of automatically generated UML diagrams (Figures \ref{fig:srv6_manager_pb2}, \ref{fig:srv6pmCommons_pb2}, \ref{fig:srv6pmReflector_pb2}, \ref{fig:srv6pmSender_pb2}).

\begin{lstlisting}[language=protobuf2, style=protobuf, caption={SRv6Manager service}]
service SRv6Manager {
  rpc Create (SRv6ManagerRequest)
    returns (SRv6ManagerReply) {}
  rpc Get (SRv6ManagerRequest)
    returns (SRv6ManagerReply) {}
  rpc Update (SRv6ManagerRequest)
    returns (SRv6ManagerReply) {}
  rpc Remove (SRv6ManagerRequest)
    returns (SRv6ManagerReply) {}
}
\end{lstlisting}

\begin{lstlisting}[language=protobuf2, style=protobuf, caption={SRv6Path}]
message SRv6Path {
  // Route of the SRv6 policy
  string destination = 1;
  // SRv6 Segment
  message SRv6Segment {
    string segment = 1;
  }
  // A collection of SRv6 Segments
  repeated SRv6Segment sr_path = 2;
  // Encap mode
  string encapmode = 3;
  // Device name
  string device = 4;
  // Table
  int32 table = 5;
}
\end{lstlisting}

\begin{lstlisting}[language=protobuf2, style=protobuf, caption={SRv6Behavior}]
message SRv6Behavior {
  // Active segment to match
  string segment = 1;
  // Action to perform
  string action = 2;
   // Nexthop
  string nexthop = 3;
  // Routing table
  int32 table = 4;
  // Interface
  string interface = 5;
  // SRv6 Segment
  message SRv6Segment {
    string segment = 1;
  }
  // A collection of SRv6 Segments
  repeated SRv6Segment segs = 6;
  // Non-loopback device
  string device = 7;
  // Local SID table
  int32 localsid_table = 8;
}
\end{lstlisting}

\begin{minipage}{\columnwidth}
\begin{lstlisting}[language=protobuf2, style=protobuf, caption={SRv6PM  service}]
service SRv6PM {
    rpc startFlowMonitoringSender (StartFlowMonitoringSenderRequest)
        returns (StartFlowMonitoringSenderReply) {}
    rpc startFlowMonitoringReflector (StartFlowMonitoringReflectorRequest)
        returns (StartFlowMonitoringReflectorReply) {}
    rpc stopFlowMonitoringSender (StopFlowMonitoringRequest)
        returns (StopFlowMonitoringReply) {}
    rpc stopFlowMonitoringReflector (StopFlowMonitoringRequest)
        returns (StopFlowMonitoringReply) {}
    rpc retriveFlowMonitoringResults (RetriveFlowMonitoringDataRequest)
        returns (FlowMonitoringDataResponse) {}
}
\end{lstlisting}
\end{minipage}

\begin{lstlisting}[language=protobuf2, style=protobuf, caption={StartFlowMonitoringSenderRequest}]
message StartFlowMonitoringSenderRequest {
    uint32 measure_id = 1;
    string sdlist = 2;
    string sdlistreverse = 3;
    repeated string in_interfaces = 4;
    repeated string out_interfaces = 5;
    SenderOptions sender_options = 6;
    ColorOptions color_options = 7;
}
\end{lstlisting}

\begin{lstlisting}[language=protobuf2, style=protobuf, caption={SenderOptions}]
message SenderOptions {
    uint32 ss_udp_port = 1;
    uint32 refl_udp_port = 2;
    MeasurementProtocol measurement_protocol = 3;
    AuthenticationMode authentication_mode = 4;
    MeasurementType measurement_type = 5;
    TimestampFormat timestamp_format = 6;
    MeasurementDelayMode measurement_delay_mode = 7;
    uint32 padding_mbz = 8;
    MeasurementLossMode measurement_loss_mode = 9;
    string authentication_key = 10;
}
\end{lstlisting}

\begin{lstlisting}[language=protobuf2, style=protobuf, caption=StartFlowMonitoringReflectorRequest]
message StartFlowMonitoringReflectorRequest {
    uint32 measure_id = 1;
    string sdlist = 2;
    string sdlistreverse = 3;
    repeated string in_interfaces = 4;
    repeated string out_interfaces = 5;
    ReflectorOptions reflector_options = 6;
    ColorOptions color_options = 7;
}
\end{lstlisting}

\begin{minipage}{\columnwidth}
\begin{lstlisting}[language=protobuf2, style=protobuf, caption=ReflectorOptions]
message ReflectorOptions {
    uint32 ss_udp_port = 1;
    uint32 refl_udp_port = 2;
    MeasurementProtocol measurement_protocol = 3;
    AuthenticationMode authentication_mode = 4;
    MeasurementType measurement_type = 5;
    MeasurementLossMode measurement_loss_mode = 6;
    string authentication_key = 7;
}
\end{lstlisting}
\end{minipage}

\FloatBarrier
\onecolumn
\section{eBPF Node Architecture}

\begin{figure*}[h]
    \centering
    \includegraphics[width=0.78\textwidth]{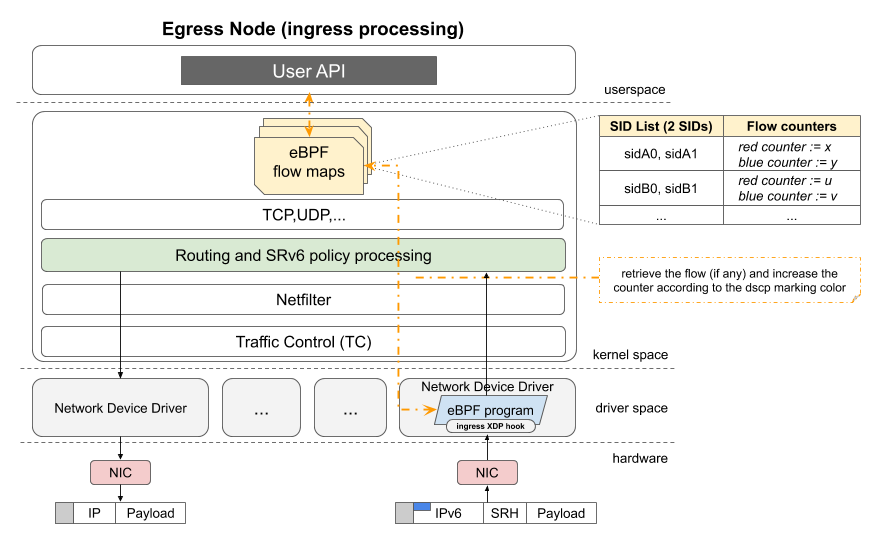}
    \caption{Packet processing and statistics management in the Egress node using the PF-PLM eBPF implementation: the figure shows how the eBPF program processes the incoming packets and how it keeps track of per-flow statistics in the egress node.}
    \label{fig:ebpfEgressNode}
    \vspace{-3ex}
\end{figure*}

\begin{figure*}[h]
    \centering
    \includegraphics[width=0.78\textwidth]{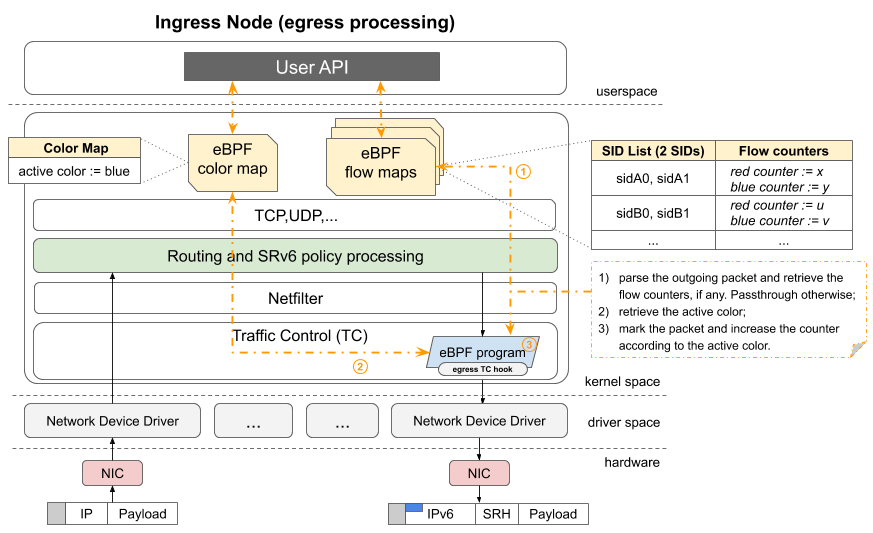}
    \caption{Packet processing and statistics management in the Ingress node using the PF-PLM eBPF implementation: the figure shows how the eBPF program processes the outgoing packets and how it keeps track of per-flow statistics in the ingress node.}
    \label{fig:ebpfIngressNode}
    \vspace{-3ex}
\end{figure*}





\end{document}